\documentstyle[12pt]{article}

\newcommand{\be}{\begin{equation}}
\newcommand{\ee}{\end{equation}}

\begin{document}
%%%%%%%%%%%%%%%%%%%%%%%%%%%%%%%%%%%%%%%%%%%%%%%%%%%%%%%%%%%%%%%%%%

\begin{center}
Annals of Physics 316 (2005) 393-413
\end{center}

\begin{center}
{\Large \bf Stationary Solutions of Liouville Equations for

\vskip 2mm
Non-Hamiltonian Systems }

\vskip 7mm
{\large \bf Vasily E. Tarasov }

{\it Skobeltsyn Institute of Nuclear Physics, \\
Moscow State University, Moscow 119992, Russia}
 
E-mail: tarasov@theory.sinp.msu.ru 
\end{center}

\begin{abstract}
We consider the class of non-Hamiltonian and dissipative statistical systems 
with distributions that are determined by the Hamiltonian.
The distributions are derived analytically as stationary
solutions of the Liouville equation for non-Hamiltonian systems. 
The class of non-Hamiltonian systems can be described by 
a non-holonomic (non-integrable) constraint: 
the velocity of the elementary phase volume change is
directly proportional to the power of non-potential forces.  
The coefficient of this proportionality is determined by Hamiltonian. 
The constant temperature systems, 
canonical-dissipative systems, and Fermi-Bose 
classical systems are the special 
cases of this class of non-Hamiltonian systems.
\end{abstract}

\vskip 3 mm


%%%%%%%%%%%%%%%%%%%%%%%%%%%%%%%%%%%%%%%%%%%%%%%%%%%%%%%%%%%%%%%%%%%%%
\section{Introduction}
%%%%%%%%%%%%%%%%%%%%%%%%%%%%%%%%%%%%%%%%%%%%%%%%%%%%%%%%%%%%%%%%%%%%%

The canonical distribution for the Hamiltonian systems was 
defined by Gibbs 
in the book "Elementary principles in statistical mechanics" \cite{Gibbs}, 
published in 1902.
In general, classical systems are not Hamiltonian systems
and the forces are the sum of potential
and non-potential forces.
Non-Hamiltonian and dissipative systems can have the same
distributions as Hamiltonian systems.
The canonical distributions for the non-Hamiltonian and dissipative 
systems were considered 
in \cite{HLM,E,EHFML,HG,Nose1,Nose2,EM,Nose,Tuck2,mplb}.

The aim of this work is the extension of the statistical
mechanics of conservative Hamiltonian systems to a
wide class of non-Hamiltonian and dissipative systems.

Let us point out non-Hamiltonian systems with 
distribution functions that are defined by the Hamiltonian.

(1) In the papers \cite{HLM,E,EHFML,HG,EM,Nose}, the constant 
temperature systems with minimal Gaussian constraint are considered.
These systems are the non-Hamiltonian systems
that are described by the non-potential forces in the form
${\bf F}^{(n)}_i=-\gamma {\bf p}_i$ and the 
Gaussian non-holonomic constraint. 
Note that this constraint can be represented 
as an addition term to the non-potential force.

(2) In the papers \cite{SET,Eb},  the canonical-dissipative systems
are considered.
These systems are the non-Hamiltonian systems
that are described by the non-potential forces 
${\bf F}^{(n)}_i=-\partial G(H)/ \partial {\bf p}_i$,
where $G(H)$ is a function of Hamiltonian $H$. 
Note that the distribution functions are derived as 
solutions the Fokker-Planck equation. 
It is known that Fokker-Planck equation can be derived
from the Liouville equation \cite{Is}.

(3) In the paper \cite{GA}, the systems with non-holonomic constraint 
and non-potential forces ${\bf F}^{(n)}_i=0$ are considered.
The equations of motion for this system are incorrect \cite{Dob}. 
The correct form of the equations 
is derived in \cite{GA} by the limit $\tau \rightarrow 0$. 
This procedure removes the incorrect term of the equations.

(4) In the paper \cite{mplb}, the canonical distribution
is considered as a stationary solution of the Liouville equation 
for a wide  class of non-Hamiltonian system.
This class is defined by a very simple condition 
for the non-potential forces:
the power of the non-potential forces must be directly proportional
to the velocity of the Gibbs phase (elementary phase volume) change.
This condition defines the general constant temperature systems.
Note that the condition is a non-holonomic constraint. This constraint 
leads to the canonical distribution as a stationary solution 
of the Liouville equations.
For the linear friction, we derived the constant temperature systems. 
The general form of the non-potential forces 
is not derived in \cite{mplb}.

(5) In the paper \cite{Tarpre02}, the quantum non-Hamiltonian systems
with pure stationary states are considered. The correspondent
classical systems are not discussed.

(6) In the paper \cite{chaos}, the non-Gaussian distributions 
are suggested for the non-Hamiltonian systems in the fractional
phase space. Note that  non-dissipative systems with the usual phase space
are dissipative systems in the fractional phase space \cite{chaos}.

Khintchin \cite{Khin} revealed the deep relation between the 
Gaussian central limit theorem and canonical Gibbs distribution. 
However, the Gaussian central limit theorem is non-unique. 
Levy and Khintchin have generalized the Gaussian central limit theorem 
to the case of summation of independent, identically distributed random 
variables which are described by long tailed distributions. 
In this case, non-Gaussian distributions replace 
the Gaussian in the generalized  limit theorems. 
It is interesting to find statistical mechanics and thermodynamics  
that is based on non-Gaussian and non-canonical distributions
\cite{AGW,Che,ZM,Barkai}. 

The aim of this paper is the description of non-Hamiltonian and 
dissipative systems with (canonical and non-canonical) 
distributions that are defined by Hamiltonian. 
This class can be described by the non-holonomic (non-integrable)
constraint: the velocity of the elementary phase volume change must
be directly proportional to the power of non-potential forces.  
The coefficient of this proportionality is determined by the Hamiltonian. 
These distributions can be derived analytically as solutions 
of the Liouville equation for non-Hamiltonian systems.
The special constraint allows us to derive solutions for 
the system, even in far-from equilibrium states. 
This class of the non-Hamiltonian systems 
is characterized by the distribution functions that are
determined by the Hamiltonian. 
The constant temperature systems \cite{HLM,E,EHFML,HG,EM,Nose}, 
the canonical-dissipative systems \cite{SET,Eb}, and the Fermi-Bose 
classical systems \cite{Eb} are the special cases 
of suggested class of non-Hamiltonian systems.

%%%%%%%%%%%%%%%%%%%%%%%%%%%%%%%%%%%%%%%%%%%%%%%%%%%%%%%%%%%%%%%%%%%%

In section 2, the definitions of the non-Hamiltonian and
dissipative systems, mathematical background and notations 
are considered.
In section 3, we consider the condition for the non-potential forces. 
We formulate the proposition that 
allows us to answer the following question:
Is this system a canonical non-Hamiltonian system?
We derive the solution of the N-particle
Liouville equation for the non-Hamiltonian systems with
non-holonomic constraint. 
In section 4, we consider the non-holonomic constraint for
non-Hamiltonian systems. 
We formulate the proposition which allows us to derive 
the canonical non-Hamiltonian systems from the equations of 
non-Hamiltonian system motion.
The non-Hamiltonian systems with the simple Hamiltonian 
and the simple non-potential forces are considered. 
In section 5, we derive the class of non-Hamiltonian systems with canonical 
Gibbs distribution as a solution of the Liouville equation. 
In section 6, we consider the non-Gaussian distributions as solutions of 
the Liouville equations for the non-Hamiltonian systems.
In section 7, we derive the analog of thermodynamics laws 
for the non-Hamiltonian systems with the distributions that are
defined by Hamiltonian.
Finally, a short conclusion is given in section 8.

%%%%%%%%%%%%%%%%%%%%%%%%%%%%%%%%%%%%%%%%%%%%%%%%%%%%%%%%%%%%%%%%%%%%%%%
\section{Definitions of Non-Hamiltonian, Dissipative and Canonical 
Non-Hamiltonian Systems}

Let us consider the definitions of non-Hamiltonian and dissipative 
classical systems \cite{Tartmf3}, which are used for the formulation 
of our results.

Usually a classical system is called a Hamiltonian system if the 
equations of motion are determined by Hamiltonian.
The more consistent definition of the non-Hamiltonian system 
is connected with Helmholtz condition for the equation of motion.\\

{\bf Definition 1.}
{\it A classical system which is defined by the equations
\be \label{EM-GF} \frac{d q_i}{dt}=G_i ,\quad
\frac{d p_i}{dt}=F_i, \ee
where $i=1,...,N$, is called Hamiltonian system if 
the right-hand sides of Eq. (\ref{EM-GF}) satisfy
the Helmholtz conditions for the phase space 
\be \label{HC} \frac{\partial G_{i}}{\partial p_j}-
\frac{\partial G_{j}}{\partial p_i}= 0, \quad
\frac{\partial F_{i}}{\partial p_j}+
\frac{\partial G_{j}}{\partial q_i}=0, \quad
\frac{\partial F_{i}}{\partial q_j}-
\frac{\partial F_{j}}{\partial q_i}= 0. \ee
Here $G_i=G_i(q,p)$, $F_i=F_i(q,p,a,t)$, 
where $a$ is a set of external parameters.}\\

If the Helmholtz conditions are satisfied, then
the equations of motion for the system (\ref{EM-GF})
can be represented as canonical equations
\be \label{EM-H} \frac{d q_i}{dt}=
\frac{\partial H}{\partial p_i} , \quad
\frac{d p_i}{dt}=-\frac{\partial H}{\partial q_i}, \ee
which are completely characterized by the Hamiltonian  
$H=H(q,p,a)$. In this case,
the forces, which act on the particles are potential forces.

%%{\bf Remark} 
If the functions $G_i$ for the non-Hamiltonian system 
(\ref{EM-GF}) are determined by the Hamiltonian 
\be \label{Gi}
G_i=\frac{\partial H}{\partial p_i},  \ee
and the Hamiltonian is a smooth function on the momentum space, 
then the first condition (\ref{HC}) is satisfied
\[ \frac{\partial^2 H}{\partial p_i \partial p_j}-
\frac{\partial^2 H}{\partial p_j \partial p_i}=0 . \]
In this case, the second condition (\ref{HC}) has the form
\be \label{HC2b} \frac{\partial F_{i}}{\partial p_j}+
\frac{\partial^2 H}{\partial q_i \partial p_j}=0 . \ee
In general, the second term does not vanish.
For example, in the nonlinear one-dimensional sigma-model \cite{Tarpl94} 
the second term of the left-hand side of Eq. (\ref{HC2b}) is 
defined by the metric.\\

{\bf Definition 2.}
{\it A mechanical system is called non-Hamiltonian if at least one 
of conditions (\ref{HC}) is not satisfied.} \\

Let us consider the time evolution of the classical state
which is defined by the distribution function 
$\rho_N(q,p,a,t)$.
The N-particle distribution function in the Hamilton picture
(for the Euler variables) is normalized by the condition
\be \label{norm} \int \rho_N(q,p,a,t) d^N q d^N p=1. \ee
The evolution equation of the distribution function
$\rho_N(q,p,a,t)$ is Liouville equation
in the Hamilton picture  
\be \label{Liu1} \frac{d\rho_N(q,p,a,t)}{dt}=
-\Omega (q,p,a,t) \rho_N(q,p,a,t). \ee
This equation describes the change of the distribution function
$\rho_N$ along the trajectory in the 6N-dimensional phase space.
Here, $\Omega$ is defined by
\be \label{Omega} \Omega(q,p,a,t)=
\frac{\partial F_i}{\partial p_i}+
\frac{\partial G_i}{\partial q_i}. \ee
Here and later we mean the sum on the repeated index
$i$ from 1 to N.
Derivative $d/dt$ is a total time derivative
\be \label{ddt} \frac{d}{dt}=\frac{\partial}{\partial t}+
G_i\frac{\partial}{\partial q_i}+
F_i\frac{\partial}{\partial p_i} . \ee
If the vector function $G_i$ is defined by Eq. (\ref{Gi}), then
\[ \Omega(q,p,a,t)= \frac{\partial F_i}{\partial p_i}+
\frac{\partial^2 H}{\partial q_i \partial p_i}. \]
In general, the second term does not vanish, for example,
in the nonlinear one-dimensional sigma-model \cite{Tarpl94}.

In the Liouville picture (for the Lagrange variables) 
the function $\Omega$ defines
the velocity of the phase volume change \cite{Tarkn1}
\[ \frac{dV_{ph}(a,t)}{dt}=\int \Omega(q,p,a,t) d^N q d^N p. \]

{\bf Definition 3.}
{\it If $\Omega\le0$ for all phase space points $(q,p)$
and $\Omega<0$ for some points of phase space, 
then the system is called a dissipative system. }\\

We can define dissipative system using 
a phase density of entropy
\[ S(q,p,a,t)=-k \ ln \ \rho_N(q,p,a,t).\]
This function usually called the Gibbs phase.
Eq. (\ref{Liu1}) leads to the equation
for the entropy density (Gibbs phase)
\be \label{dS} \frac{dS(q,p,a,t)}{dt}=
k\Omega (q,p,a,t) .\ee
It is easy to see that the function $\Omega$ is proportional to the
velocity of the phase entropy density change.
Therefore, the dissipative systems can be defined by the following
equivalent definition. \\

{\bf Definition 4.}
{\it A system is called a generalized dissipative system if the velocity 
of the entropy density change does not equal to zero. } \\

Let us define the special class of the non-Hamiltonian systems with 
distribution functions that are completely characterized by the Hamiltonian. 
These distributions can be derived analytically as stationary
solutions of the Liouville equation for the non-Hamiltonian system. \\

{\bf Definition 5.}
{\it A non-Hamiltonian system will be called 
a canonical non-Hamiltonian system 
%%% "generalized canonical dissipative systems"
if the distribution function is determined by the Hamiltonian, 
i.e., $\rho_N(q,p,a)$ can be written in the form
\be \label{rhoH} \rho_N(q,p,a)=\rho_N(H(q,p,a), a), \ee
where $a$ is a set of external parameters.}\\

Examples of the canonical  non-Hamiltonian systems:\\
(1) The constant temperature systems \cite{HLM,E,EHFML,HG,EM,Nose}
that have the canonical distribution. In general,
these systems can be defined by the non-holonomic 
constraint, which is suggested in \cite{mplb}. \\
(2) The Fermi-Bose canonical-dissipative systems \cite{Eb} 
which are defined by the distribution functions in the form
\be \label{FermiBose}
\rho_N(H(q,p,a))=\frac{1}{exp\beta (H(q,p,a)-\mu)+s} .
\ee
(3) The classical system with the Breit-Wigner 
distribution function is defined by
\be \label{BreitWigner}
\rho_N(H)=\frac{\lambda}{(H-E)^2+(\Gamma/2)^2} .
\ee

%%%%%%%%%%%%%%%%%%%%%%%%%%%%%%%%%%%%%%%%%%%%%%%%%%%%%%%%%%%%%%%%%%
\section{Distribution as a Solution of the Liouville Equation}

%%%%%%%%%%%%%%%%%%%%%%%%%%%%%%%%%%%%%%%%%%%%
\subsection{Formulation of the Results}

Let us formulate the proposition that 
allows us to answer the following question:
Is this system a canonical non-Hamiltonian system?

Let us consider the N-particle non-Hamiltonian systems 
which are defined by the equations
\be \label{Sys}
\frac{d{\bf r}_i}{dt}=\frac{\partial H}{\partial {\bf p}_i},
\quad
\frac{d{\bf p}_i}{dt}=-\frac{\partial H}{\partial {\bf r}_i}+
{\bf F}^{(n)}_i . \ee
The power of non-potential forces is defined by
\be \label{power} {\cal P}({\bf r},{\bf p},a)=
{\bf F}^{(n)}_i \frac{\partial H}{\partial {\bf p}_i}. \ee
If the power of the non-potential forces is equal to zero 
(${\cal P}=0$) and $\partial H/ \partial t=0$, then classical system
is called a conservative system. 
The velocity of an elementary phase volume change $\Omega$ is defined 
by the equation
\be \label{omega} \Omega({\bf r},{\bf p},a)=
\frac{\partial {\bf F}_i}{\partial {\bf p}_i}+
\frac{\partial^2 H}{\partial {\bf r}_i \partial {\bf p}_i}
=\frac{\partial {\bf F}^{(n)}_i}{\partial {\bf p}_i} . \ee
We use the following notations for the scalar product
\[ \frac{\partial {\bf A}_{i}}{\partial {\bf a}_i}=
\sum^N_{i=1}\Bigl(
\frac{\partial A_{xi}}{\partial a_{xi}}+
\frac{\partial A_{yi}}{\partial a_{yi}}+
\frac{\partial A_{zi}}{\partial a_{zi}}\Bigr). \]

The aim of this section is to prove the following result. \\

{\bf Proposition 1.}
{\it If the non-potential forces ${\bf F}^{(n)}_i$ of 
the non-Hamiltonian system  (\ref{Sys}) 
satisfy the constraint condition
\be \label{NC-P1} 
g(H) {\bf F}^{(n)}_i \frac{\partial H}{\partial {\bf p}_i} -
\frac{\partial {\bf F}^{(n)}_i}{\partial {\bf p}_i}=0 , \ee
then this system is a canonical non-Hamiltonian system 
with the distribution function
\be 
\rho_N({\bf r},{\bf p},a)=Z(a) exp [ - L(H({\bf r},{\bf p},a))] ,
\ee
where the function $L(H)$ is defined by the equation
\be  g(H)=\frac{\partial L(H)}{\partial H} . \ee
}  \\

The condition (\ref{NC-P1}) can be formulated in other words:  \\
{\it If velocity of the elementary phase volume change 
$\Omega$ is directly proportional to the power ${\cal P}$ 
of non-potential forces ${\bf F}^{(n)}_i$ of the non-Hamiltonian 
system (\ref{Sys}) and coefficient of this proportionality
is a function $g(H)$ of Hamiltonian $H$, i.e., 
\be \label{NC-P0} \Omega({\bf r},{\bf p},t)-
g(H) {\cal P}({\bf r},{\bf p},t)=0 , \ee
then this system is a canonical non-Hamiltonian system.} \\

Note that any non-Hamiltonian system with the non-holonomic constraint 
(\ref{NC-P0}) or (\ref{NC-P1}) is a canonical non-Hamiltonian system. \\

{\bf Example.}
Let us consider $g(H)=3N \beta(a)$, where $\beta(a)=1/kT(a)$. 
This case is considered in \cite{mplb}.
If we consider the N-particle system with the Hamiltonian
\be \label{ex1}
H({\bf r},{\bf p},a)=\sum^N_{i=1}\frac{{\bf p}^2_i}{2m}+U({\bf r},a) ,
\ee
and a linear friction, which is defined by the non-potential forces
\be \label{ex2} {\bf F}^{(n)}_i=-\gamma {\bf p}_i, \ee
then the non-holonomic constraint (\ref{NC-P1}) has the form
\be \label{con1} \sum^N_{i=1} \frac{{\bf p}^{2}_i}{m}=kT(a), \ee
i.e., the kinetic energy of the system must be a constant. 
The constraint (\ref{con1}) is a non-holonomic minimal Gaussian
constraint \cite{mplb,Nose}. 

If the function $g(H)$ is defined by $g(H)=3N \beta(a)$, 
then the non-Hamiltonian system
can have the canonical Gibbs distribution \cite{mplb}.
The classical systems that are defined by Eqs. (\ref{ex1}) - (\ref{con1}) 
are canonical non-Hamiltonian systems.

%%%%%%%%%%%%%%%%%%%%%%%%%%%%%%%%%%%%%%%%%%%%%%%%%%%%%%%%%%%%%%%%
\subsection{Proof of the Result}
%%%{Distribution as a solution of the Liouville Equation}

Solving the Liouville equation with the non-holonomic constraint 
(\ref{NC-P1}), we can obtain the (canonical and non-canonical) 
distributions that are defined by the Hamiltonian.

Let us consider the Liouville equation for the N-particle
distribution function $\rho_N=\rho_{N}({\bf r},{\bf p},a,t)$. 
This distribution function $\rho_N$ 
express a probability that a phase space point $({\bf r},{\bf p})$ 
will appear. The Liouville equation for this non-Hamiltonian system
\be \label{rhoN} \frac{\partial \rho_N}{\partial t}+
\frac{\partial}{\partial {\bf r}_i}\Bigl({\bf G}_i \rho_N \Bigr)+
\frac{\partial}{\partial {\bf p}_i} \Bigl({\bf F}_i \rho_N \Bigr)=0\ee
expresses the conservation of probability in the phase space. 
Here, we use
\[ {\bf G}_i=\frac{\partial H}{\partial {\bf p}_i} , \quad
{\bf F}_i=-\frac{\partial H}{\partial {\bf r}_i}
+{\bf F}^{(n)}_i .\]
We define a total time derivative along the phase space trajectory by
\be \frac{d}{dt}=
\frac{\partial}{\partial t}+
{\bf G}_i \frac{\partial}{\partial {\bf r}_i}+
{\bf F}_i \frac{\partial}{\partial {\bf p}_i} . \ee
Therefore Eq. (\ref{rhoN}) can be written in the form (\ref{Liu1})
\be \label{39}
\frac{d\rho_N}{dt}=-\Omega \rho_N ,
\ee
where the omega function is defined by Eq. (\ref{omega}). 
In classical mechanics of Hamiltonian systems the right-hand side of 
the Liouville equation (\ref{39}) is zero, and 
the distribution function does not change in time.  
For the non-Hamiltonian systems (\ref{Sys}), 
the omega function (\ref{omega}) does not vanish.
For this system, the omega function is defined by Eq. (\ref{omega}). 
For the canonical non-Hamiltonian systems, this function 
is defined by the constraint (\ref{NC-P1}) in the form
\[ \Omega=g(H) {\bf F}^{(n)}_i \frac{\partial H}{\partial {\bf p}_i}. \]
In this case, the Liouville equation has the form
\be \label{41}
\frac{d\rho_N}{dt}=-
g(H) {\bf F}^{(n)}_i \frac{\partial H}{\partial {\bf p}_i} \rho_N.
\ee
Let us consider the total time derivative of the Hamiltonian.  
Using equations of motion (\ref{Sys}), we have 
\[ \frac{dH}{dt}=\frac{\partial H}{\partial t}+
\frac{\partial H}{\partial {\bf p}_i}
\frac{\partial H}{\partial {\bf r}_i}+
\Bigl(-\frac{\partial H}{\partial {\bf r}_i}
+{\bf F}^{(n)}_i \Bigr)\frac{\partial H}{\partial {\bf p}_i} = 
 \frac{\partial H}{\partial t}+
{\bf F}^{(n)}_i \frac{\partial H}{\partial {\bf p}_i} . \]
If ${\partial H}/{\partial t}=0$, then 
the power ${\cal P}$ of non-potential forces is equal to 
the total time derivative of the Hamiltonian
\[ {\bf F}^{(n)}_i \frac{\partial H}{\partial {\bf p}_i}=
\frac{dH}{dt} . \]
Eq. (\ref{41}) can be written in the form
\be \frac{d\rho_N}{dt}=-g(H) \frac{dH}{dt} \rho_N. \ee
Let us consider the following form of this equation:
\be \label{Eq45} \frac{d ln \ \rho_N}{dt}=-g(H) \frac{dH}{dt}. \ee
If $g(H)$ is an integrable function, then this function 
can be represented as a derivative 
\be  \label{L} g(H)=\frac{\partial L(H)}{\partial H} . \ee
In this case, we can write Eq. (\ref{Eq45}) in the form
\be  \frac{d ln \ \rho_N}{dt}=- \frac{dL(H)}{dt} . \ee
As a result, we have the following solution of the Liouville equation: 
\be \label{solution}
\rho_N({\bf r},{\bf p},a)=Z(a) exp - L(H({\bf r},{\bf p},a)) .
\ee
The function $Z(a)$ is defined by the normalization condition. 
It is easy to see that the distribution function of the 
non-Hamiltonian system is determined by the Hamiltonian.
Therefore, this system is a canonical non-Hamiltonian system.

Note that $N$ is an arbitrary natural number since we do not
use the condition $N>>1$ or $N \rightarrow \infty$.

%%%%%%%%%%%%%%%%%%%%%%%%%%%%%%%%%%%%%%%%%%%%%%%%%%%%%%%%%%%%%%%%%
\section{Non-Holonomic Constraint for Non-Hamiltonian Systems}

%%%%%%%%%%%%%%%%%%%%%%%%%%%%%%%%%%%%%%%%
\subsection{Formulation of the Result}

Let us formulate the proposition which allows us to derive 
the canonical non-Hamiltonian systems from any equations of 
motion of non-Hamiltonian systems.

The aim of this section is to prove the following result. \\
 
{\bf Proposition 2.}
{\it For any non-Hamiltonian system which is defined by the equation
\be \frac{d{\bf r}_i}{dt}=\frac{\partial H}{\partial {\bf p}_i},\quad
\frac{d{\bf p}_i}{dt}={\bf F}_i, \ee
where ${\bf F}_i$ is the sum of potential and 
non-potential forces 
\be \label{F} {\bf F}_i=-\frac{\partial H}{\partial {\bf r}_i}
+{\bf F}^{(n)}_i , \ee
there exists a canonical non-Hamiltonian system 
that is defined by the equations
\be \frac{d{\bf r}_i}{dt}=\frac{\partial H}{\partial {\bf p}_i},\quad
\frac{d{\bf p}_i}{dt}={\bf F}^{new}_i , \ee
where the non-potential forces ${\bf F}^{new}_i$ are defined by
\be \label{new2}
{\bf F}^{new}_i=
\frac{ {\bf A}_k{\bf A}_k \delta_{ij}-{\bf A}_i{\bf A}_j }{
{\bf A}_k{\bf A}_k} {\bf F}_j
-\frac{{\bf A}_i {\bf B}_j }{{\bf A}_k{\bf A}_k} 
\frac{\partial H}{\partial {\bf p}_j} . \ee
The vectors ${\bf A}_i$ and ${\bf B}_i$ are defined by the equations
\[ {\bf A}_i=\frac{\partial g(H)}{\partial H} 
\frac{\partial H}{\partial {\bf p}_i} 
\frac{\partial H}{\partial {\bf p}_j}{\bf F}^{(n)}_j 
+g(H)\frac{\partial {\bf F}^{(n)}_j}{\partial {\bf p}_i} 
\frac{\partial H}{\partial {\bf p}_j} + \]
\be \label{51} +g(H){\bf F}^{(n)}_j\frac{\partial^2 H}{\partial 
{\bf p}_i\partial {\bf p}_j}
-\frac{\partial^2 {\bf F}^{(n)}_j}{\partial {\bf p}_i\partial {\bf p}_j},
\ee
and
\[ {\bf B}_i=\frac{\partial g(H)}{\partial H}
\frac{\partial H}{\partial {\bf r}_i}
\frac{\partial H}{\partial {\bf p}_j}{\bf F}^{(n)}_j+ 
g(H)\frac{\partial {\bf F}^{(n)}_j}{\partial {\bf r}_i} 
\frac{\partial H}{\partial {\bf p}_j} + \]
\be \label{52}
+g(H){\bf F}^{(n)}_j\frac{\partial^2 H}{\partial 
{\bf r}_i\partial {\bf p}_j}
-\frac{\partial^2 {\bf F}^{(n)}_j}{\partial {\bf r}_i\partial {\bf p}_j}
 . \ee
}

Note that the forces that are defined by Eqs. (\ref{new2}) - (\ref{52}) 
satisfy the non-holonomic constraint (\ref{NC-P0}), i.e.,
\be \label{NC-new} 
g(H) {\bf F}^{new}_j \frac{\partial H}{\partial {\bf p}_j} -
\frac{\partial {\bf F}^{new}_j}{\partial {\bf p}_j} 
-\frac{\partial^2 H}{\partial {\bf r}_j \partial {\bf p}_j}=0. \ee

%%%%%%%%%%%%%%%%%%%%%%%%%%%%%%%%%%%%%%%%%%%%%%%%%%%%%%%%
\subsection{Proof. Part I}

In this subsection, we prove Eq. (\ref{new2}).

Let us consider the N-particle classical system in 
the Hamilton picture.
Denote the position of the $i$th particle 
by ${\bf r}_i$ and its momentum by ${\bf p}_i$. 

Suppose that the system is subjected to a non-holonomic 
(non-integrable) constraint in the form 
\be \label{NC} f({\bf r},{\bf p})=0 . \ee
Differentiation of Eq. (\ref{NC}) with respect to time gives a relation
\be \label{TD}
{\bf A}_i({\bf r},{\bf p}) \frac{d{\bf p_i}}{dt}+
{\bf B}_i({\bf r},{\bf p})\frac{d{\bf r_i}}{dt}=0,
\ee
where 
\be \label{AB} 
{\bf A}_i({\bf r},{\bf p})=\frac{\partial f}{\partial {\bf p}_i}, \quad
{\bf B}_i({\bf r},{\bf p})=\frac{\partial f}{\partial {\bf r}_i}.
\ee
An unconstrained motion of the $i$th particle,
where $i=1,...,N$, is described by the equations
\be \label{EM1}\frac{d{\bf r}_i}{dt}={\bf G}_i , \quad
\frac{d{\bf p}_i}{dt}={\bf F}_i,\ee
where ${\bf F}_i$ is a resulting force, which acts on the $i$th particle.

The unconstrained motion gives a trajectory which leaves the constraint 
hypersurface (\ref{NC}).
The constraint forces ${\bf R}_i$ must be added to the equation 
of motion to prevent the deviation from the constraint hypersurface
\be \label{EM2} \frac{d{\bf r}_i}{dt}={\bf G}_i , \quad
\frac{d{\bf p}_i}{dt}={\bf F}_i+{\bf R}_i .\ee
The constraint force ${\bf R}_i$ for the non-holonomic 
constraint is proportional to the ${\bf A}_i$ \cite{Dob}
\be
{\bf R}_{i}=\lambda {\bf A}_i ,
\ee
where the coefficient $\lambda$ of the constraint force term 
is an undetermined Lagrangian multiplier.
For the non-holonomic constraint (\ref{NC}),  
equations of motion (\ref{EM1}) are modified as
\be \label{EM3} \frac{d{\bf r}_i}{dt}={\bf G}_i , \quad
\frac{d{\bf p}_i}{dt}={\bf F}_i+\lambda{\bf A}_i .\ee

The Lagrangian coefficient $\lambda$ is determined 
by Eq. (\ref{TD}).
Substituting Eq. (\ref{EM2}) into Eq. (\ref{TD}), we get
\be \label{TD2}
{\bf A}_i ({\bf F}_i+\lambda{\bf A}_i)+
{\bf B}_i {\bf G}_i=0 . \ee
Therefore, the Lagrange multiplier $\lambda$ is equal to
\be \label{TD5}
\lambda =-\frac{{\bf A}_i {\bf F}_i
+{\bf B}_i {\bf G}_i }{{\bf A}_k{\bf A}_k } .
\ee
As a result, we obtain the following equations:
\be \label{EM4} \frac{d{\bf r}_i}{dt}={\bf G}_i, \quad
\frac{d{\bf p}_i}{dt}={\bf F}_i
-{\bf A}_i\frac{{\bf A}_j {\bf F}_j
+{\bf B}_j  {\bf G}_j }{ {\bf A}_k{\bf A}_k}.\ee
These equations we can rewrite in the form (\ref{EM1}) 
\be \label{EM6} \frac{d{\bf r}_i}{dt}={\bf G}_i, \quad
\frac{d{\bf p}_i}{dt}={\bf F}^{new}_i \ee
with the new forces
\be \label{new}
{\bf F}^{new}_i=
\frac{ {\bf A}_k{\bf A}_k \delta_{ij}-{\bf A}_i{\bf A}_j }{
{\bf A}_k{\bf A}_k} {\bf F}_j
-\frac{{\bf A}_i {\bf B}_j }{{\bf A}_k{\bf A}_k} {\bf G}_j . \ee
In general, the forces ${\bf F}^{new}_i$ are non-potentials forces 
(see examples in \cite{mplb}).

Eqs. (\ref{EM4}) are equations of 
the {\it holonomic} non-Hamiltonian system. For any trajectory 
of the system in the phase space, we have $f=const$.
If initial values ${\bf r}_k(0)$ and ${\bf p}_k(0)$ satisfy 
the constraint condition $f({\bf r}(0),{\bf p}(0))=0$,
then solution of Eqs. (\ref{EM4}) and (\ref{new}) 
is a motion of the non-holonomic system.

%%%%%%%%%%%%%%%%%%%%%%%%%%%%%%%%%%%%%%%%%%%%%%%%%%%%%%%%%%
\subsection{Proof. Part II.}

In this subsection we prove Eqs. (\ref{51}) and (\ref{52}). 

Let us consider the non-Hamiltonian system (\ref{EM1}) with
\be
{\bf G}_i=\frac{\partial H}{\partial {\bf p}_i}, \quad
{\bf F}_i=-\frac{\partial H}{\partial {\bf r}_i}+{\bf F}^{(n)}_i, 
\ee
and the special form of the non-holonomic constraint (\ref{NC}).
Let us assume the following constraint: 
the velocity of the elementary phase volume change 
\ $\Omega({\bf r},{\bf p},a)$
is directly proportional to the power 
${\cal P}({\bf r},{\bf p},a)$
of the non-potential forces, i.e.,  
\be \label{NC2} \Omega({\bf r},{\bf p},a)=
g(H) {\cal P}({\bf r},{\bf p},a), \ee
where $g(H)$ depends on the Hamiltonian $H$.
Therefore, the system is subjected to a non-holonomic (non-integrable)
constraint (\ref{NC}) in the form 
\be \label{fPO} f({\bf r},{\bf p},a)= g(H) {\cal P}({\bf r},{\bf p},a)
-\Omega({\bf r},{\bf p},a)=0 . \ee
This constraint is a generalization of the condition 
which is suggested in \cite{mplb}. 
The power ${\cal P}$ of the 
non-potential forces ${\bf F}^{(n)}_i$ is defined by Eq. (\ref{power}). 
The function $\Omega$ is defined by Eq. (\ref{Omega}).

Eq. (\ref{fPO}) for the non-potential forces has the form  
\[ g(H) {\bf F}^{(n)}_j \frac{\partial H}{\partial {\bf p}_j} -
\frac{\partial {\bf F}^{(n)}_j}{\partial {\bf p}_j}=0. \]

Let us find the functions ${\bf A}_i$ and ${\bf B}_i$ for this constraint.
Differentiation of the function $f({\bf r},{\bf p},a)$ 
with respect to ${\bf p}_i$ gives
\[ {\bf A}_i=\frac{\partial f}{\partial {\bf p}_i} = 
\frac{\partial}{\partial {\bf p}_i} \Bigl(
g(H) {\bf F}^{(n)}_j \frac{\partial H}{\partial {\bf p}_j}\Bigr) 
-\frac{\partial}{\partial {\bf p}_i}
\frac{\partial {\bf F}^{(n)}_j}{\partial {\bf p}_j}  . \]
Therefore, we obviously have (\ref{51}).
Differentiation of the function $f({\bf r},{\bf p},t)$ 
with respect to ${\bf r}_i$ gives
\[ {\bf B}_i=\frac{\partial f}{\partial {\bf r}_i} = 
\frac{\partial}{\partial {\bf r}_i} \Bigl(
g(H) {\bf F}^{(n)}_j \frac{\partial H}{\partial {\bf p}_j}\Bigr) 
-\frac{\partial}{\partial {\bf r}_i}
\frac{\partial {\bf F}^{(n)}_j}{\partial {\bf p}_j} . \]
Therefore, we have (\ref{52}).

%%%%%%%%%%%%%%%%%%%%%%%%%%%%%%%%%%%%%%%%%%%%%%%%%%%%%%%%%%%%%%%%%%
\subsection{Minimal Constraint Models}

To realize simulation of the classical systems with 
canonical and non-canonical distributions, we must have 
the simple constraints. 
Let us consider the minimal constraint models which
are defined by the simplest form of the Hamiltonian and
the non-potential forces
\be
H({\bf r},{\bf p},a)=\frac{{\bf p}^2}{2m}+U({\bf r},a) ,
\quad {\bf F}^{(n)}_i=-\gamma {\bf p}_i .
\ee
where ${\bf p}^2=\sum^N_{i=1}{\bf p}^2_i$.
For these models, the non-holonomic constraint 
is defined by the equation
\be
f=g(H)\frac{{\bf p}^2}{m}-3N=0 ,
\ee
where $N$ is the number of particles. 
The phase space gradients (\ref{51}) and (\ref{52}) 
of the constraint can be represented in the form
\[ {\bf A}_i=\Bigl(\frac{\partial g(H)}{\partial H}
\frac{{\bf p}^2}{2m}+g(H)\Bigr)\frac{2{\bf p}_i}{m} ,
\quad
{\bf B}_i=\frac{\partial g(H)}{\partial H}
\frac{\partial H}{\partial {\bf r}_i} . \]

The non-potential forces of the minimal constraint models
have the form
\[ {\bf F}_i=-
\frac{{\bf p}^2\delta_{ij}-{\bf p}_i{\bf p}_j}{{\bf p}^2}
\frac{\partial U}{\partial {\bf r}_j}
+\frac{{\bf p}_i{\bf p}_j }{2{\bf p}^2
(({\bf p}^2/2m) ( \partial g(H)/ \partial H)+g(H))} 
\frac{\partial g(H)}{\partial H} \frac{\partial U}{\partial {\bf r}_j} . \]
It is easy to see that all minimal constraint models 
have the potential forces. 

For the minimal Gaussian constraint model 
\[ \frac{\partial g(H)}{\partial H}=0, \]
we have the non-potential forces in the form
\[ {\bf F}_i=-\frac{\partial U}{\partial {\bf r}_j}
\frac{{\bf p}^2\delta_{ij}-{\bf p}_i{\bf p}_j}{{\bf p}^2} . \]
This model describes the constant temperature systems
\cite{HLM,E,EHFML,HG,EM,Nose,mplb}.

%%%%%%%%%%%%%%%%%%%%%%%%%%%%%%%%%%%%%%%%%%%%%%%%%%%%%%%%%%%%%%%%%%
\subsection{Minimal Gaussian Constraint Model}

Let us consider the N-particle system with the Hamiltonian
\be
H({\bf r},{\bf p},a)=\frac{{\bf p}^2}{2m}+U({\bf r},a) ,
\ee
the function $g(H)=3N/kT$, and the linear friction 
\be \label{fric} {\bf F}^{(n)}_i=-\gamma {\bf p}_i, \ee
where $i=1,...,N$. 
Note that $N$ is an arbitrary natural number.
Substituting Eq. (\ref{fric}) into Eqs. (\ref{power}) 
and (\ref{omega}), we get
the power ${\cal P}$ and the omega function $\Omega$:
\[ {\cal P}=-\frac{\gamma}{m} {\bf p}^2, \quad \Omega=-3\gamma N. \]
The non-holonomic constraint has the form
\be \label{PO2} \frac{{\bf p}^{2}}{m}=kT(a), \ee
i.e., the kinetic energy of the system must be a constant.
Note that Eq. (\ref{PO2}) has not the friction parameter $\gamma$. 

For the N-particle system with friction (\ref{fric})
and non-holonomic constraint (\ref{PO2}), we have 
the following equations of motion 
\be \label{em} \frac{d{\bf r}_i}{dt}=
\frac{{\bf p}_i}{m} , \quad
\frac{d{\bf p}_i}{dt}=-\frac{\partial U}{\partial {\bf r}_i}
-\gamma {\bf p}_i+
\lambda \frac{\partial f}{\partial {\bf p}_i}, \ee
where the function $f$ is defined by
\be \label{con} f({\bf r},{\bf p})=
\frac{1}{2}\Bigl({\bf p}^{2}-mkT \Bigr):
\quad f({\bf r},{\bf p})=0. \ee
Eq. (\ref{em}) and condition (\ref{con})
define 6N+1 variables $({\bf r},{\bf p},\lambda)$.

Let us find the Lagrange multiplier $\lambda$.
Substituting Eq. (\ref{con}) into Eq. (\ref{em}), we get
\be \label{em2}
\frac{d{\bf p}_i}{dt}=-\frac{\partial U}{\partial {\bf r}_i}
+(\lambda-\gamma) {\bf p}_i . \ee
Using $df/dt=0$ in the form 
\be \label{pp0} {\bf p}_i \frac{d{\bf p}_i}{dt}=0 \ee
and substituting  Eq. (\ref{em2}) into 
Eq. (\ref{pp0}), we get
the Lagrange multiplier $\lambda$ in the form
\[ \lambda= \frac{1}{mkT}
{\bf p}_j\frac{\partial U}{\partial {\bf r}_j}+\gamma . \]
As a result, we have the holonomic system 
that is defined by the equations
\be \label{em4} \frac{d{\bf r}_i}{dt}=\frac{{\bf p}_i}{m} , \quad
\frac{d{\bf p}_i}{dt}= \frac{1}{mkT}
{\bf p}_i {\bf p}_j \frac{\partial U}{\partial {\bf r}_j}
-\frac{\partial U}{\partial {\bf r}_i}. \ee
This system is equivalent to the non-holonomic system (\ref{em}). 
For the classical N-particle system (\ref{em4}), 
condition (\ref{PO2}) is satisfied.
If the time evolution of the N-particle system 
is defined by Eq. (\ref{em4}) or Eqs. (\ref{em}) and (\ref{con}), 
then we have the canonical distribution function in the form
\be \label{cdf} \rho({\bf r},{\bf p},a,T)=exp\frac{1}{kT}
\Bigl({\cal F}(a,T)-H({\bf r},{\bf p},a)\Bigr). \ee
For example, the N-particle system with the forces
\be {\bf F}_i= \frac{\omega^2(a)}{kT} {\bf p}_i
{\bf p}_j{\bf r}_j-m\omega^2(a) {\bf r}_i \ee
can have canonical distribution (\ref{cdf}) of 
the linear harmonic oscillator with
\[ U({\bf r},a)=\frac{m\omega^2(a) {\bf r}^2}{2}. \]

%%%%%%%%%%%%%%%%%%%%%%%%%%%%%%%%%%%%%%%%%%%%%%%%%%%%%%%%%%%%%%%%%%%
\section{Canonical Distributions}

In this section, we consider the subclass of the canonical 
non-Hamiltonian system that is described by canonical distribution.  
This subclass of the canonical non-Hamiltonian N-particle
system is defined by the simple function $g(H)=3N \beta(a)$ 
in the non-holonomic constraint (\ref{NC-P0}). \\

{\bf Proposition 3.}
{\it If velocity of the elementary phase volume change 
is directly proportional to the power of non-potential forces, 
then we have the usual canonical Gibbs distribution as a 
solution of the Liouville equation. } \\

In other words, the non-Hamiltonian system with the
non-holonomic constraint
\be \label{PO} \Omega=\beta(a) {\cal P} \ee
can have the canonical Gibbs distribution 
\[ \rho_N=exp \beta(a) \Bigl(
{\cal F}(a)- H({\bf r},{\bf p},a) \Bigr) \]
as a solution of the Liouville equation.
Here, the coefficient $\beta(a)$ does not depend
on $({\bf r},{\bf p},t)$, i.e.,  
\[ d\beta(a)/dt=0 . \]

For the non-Hamiltonian systems, 
the omega function (\ref{omega}) does not vanish.
Using Eq. (\ref{power}), we have
\be
\Omega=\beta(a) {\bf F}^{(n)}_i \frac{\partial H}{\partial {\bf p}_i}.
\ee
In this case, the Liouville equation has the form
\be \label{91}
\frac{d\rho_N}{dt}=-
\beta(a) {\bf F}^{(n)}_i \frac{\partial H}{\partial {\bf p}_i} \rho_N.
\ee
Let us consider the total time derivative for the Hamiltonian
\be \label{dH} \frac{dH}{dt}=\frac{\partial H}{\partial t}+
{\bf F}^{(n)}_i \frac{\partial H}{\partial {\bf p}_i} . \ee
If ${\partial H}/{\partial t}=0$, then the energy change 
is equal to the power ${\cal P}$ of the
non-potential forces ${\bf F}^{(n)}_i$.  
Eq. (\ref{91}) can be written in the form
\be \frac{d\rho_N}{dt}=-\beta(a) \frac{dH}{dt} \rho_N. \ee
Therefore, the Liouville equation can be rewritten in the form
\[ \frac{d \ ln \rho_N({\bf r},{\bf p},a,t)}{dt}+
\beta(a)\frac{dH({\bf r},{\bf p},a)}{dt}=0. \]
Since coefficient $\beta(a)$ is a constant ($d\beta(a)/dt=0$), we have
\[  \frac{d}{dt}\Bigl( ln \rho_N({\bf r},{\bf p},a,t)+
\beta(a) H({\bf r},{\bf p},a) \Bigr)=0, \]
i.e., the value $(ln \rho_N+\beta H)$ is a constant along the trajectory
of the system in 6N-dimensional phase space.
Let us denote this constant value by $\beta(a) {\cal F}(a)$.
Then, we have
\[ ln \rho_N({\bf r},{\bf p},a,t)+
\beta(a) H({\bf r},{\bf p},a)=\beta (a){\cal F}(a),\]
where $d{\cal F}(a)/dt=0$. It follows that
\[ ln \ \rho_N({\bf r},{\bf p},a,t)=\beta(a)\Bigl(
{\cal F}(a)- H({\bf r},{\bf p},a) \Bigr) . \]
As a result, we have a canonical distribution function
\[ \rho_N({\bf r},{\bf p},a,t)=exp \beta(a) \Bigl(
{\cal F}(a)- H({\bf r},{\bf p},a) \Bigr) \]
in the Hamilton picture. The value ${\cal F}(a)$ is defined
by the normalization condition (\ref{NC}).

Therefore the distribution of this non-Hamiltonian system is
a canonical distribution.

Note that $N$ is an arbitrary natural number since we do not
use the condition $N>>1$ or $N \rightarrow \infty$.

%%%%%%%%%%%%%%%%%%%%%%%%%%%%%%%%%%%%%%%%%%%%%%%%%%%%%%%%%%%%%%%%%%%
\section{Non-Canonical Distributions}

The well-known non-Gaussian distribution is the Breit-Wigner distribution.
This distribution has a probability density function in the form 
\be \rho(x)=\frac{1}{\pi(1+x^2)} . \ee
The Breit-Wigner distribution is also known in statistics 
as Cauchy distribution.  
The Breit-Wigner distribution is a generalized form 
originally introduced \cite{Breit36} to describe 
the cross-section of resonant nuclear scattering in the form
\be 
\rho(H)=\frac{\lambda}{(H-E)^2+(\Gamma/2)^2} .
\ee
This distribution can be
derived from the transition probability of a resonant 
state with known lifetime \cite{Bohr69,Fermi51,Paul69}.

If the function $g(H)$ of the non-holonomic constrain
is defined by
\be
g(H)=\frac{2(H-E)}{(H-E)^2+(\Gamma/2)^2} ,
\ee
then we have non-Hamiltonian systems with the Breit-Wigner 
distribution as a solution of the Liouville equation.

If the function $g(H)$ of the non-holonomic constrain has the form
\be
g(H)=\frac{\beta(a)}{1+\alpha\  exp \ \beta(a) H},
\ee
then we have classical non-Hamiltonian systems with Fermi-Bose 
distribution (\ref{FermiBose}) considered by Ebeling \cite{Eb}. 
This distribution can be derived 
as a solution of the Liouville equation.
Note that Ebeling derives the Fermi-Bose distribution function as 
a solution of the Fokker-Planck equation. 
It is known that Fokker-Planck equation can be derived
from the Liouville equation \cite{Is}.

If the non-potential forces ${\bf F}^{(n)}_i$ are determined 
by the Hamiltonian 
\be {\bf F}^{(n)}_i=-\partial G(H)/ \partial {\bf p}_i , \ee
then we have the canonical non-Hamiltonian systems,
which are considered in \cite{SET,Eb}. These systems 
are called canonical dissipative systems.

Note that the linear function $g(H)$ in the form
\[ g(H)=\beta_1(a)+\beta_2(a)H \]
leads to the following non-canonical distribution function
\be
\rho_N=Z(a) exp-\Bigl( \beta_1(a)H+\frac{1}{2}\beta_2(a)H^2 \Bigr) .
\ee
The proof of this proposition can be directly derived from 
Eqs. (\ref{solution}) and (\ref{L}).

Let us assume that Eq. (\ref{rhoH}) can be solved in the form 
\be H=\theta (a) h(\rho_N) , \ee
where $h$ depends on the distribution $\rho_N$. The function
$\theta(a)$ is a function of the parameters $a$.  
In this case, the function $g(H)$ is a composite function
\be
R(\rho_N)=-g(\theta(a)h(\rho_N)). 
\ee
This function can be defined by
\be
R(\rho_N)=\frac{1}{\rho_N} \left(\theta(a)
\frac{\partial h(\rho_N)}{\partial \rho_N} \right)^{-1}.
\ee
In this case, the Liouville equation for the 
non-Hamiltonian system has the form
\be
\frac{d\rho_N}{dt}=R(\rho_N) {\cal P} .
\ee
This equation is a nonlinear equation.
Note that the classical Fermi-Bose systems 
\cite{Eb} have the function in the form
\be
R(\rho_N)=-\beta(a)(\rho_N-s \rho^2_N) .
\ee

The nonlinearity of the Liouville equation is not connected with 
an incorrectly defined phase space. 
This nonlinearity is a symptom of the use of an 
incorrectly defined boundary condition. The Bogoliubov principle of 
correlation weakening cannot be used for classical Fermi-Bose systems. 
The classical Fermi-Bose systems can be considered as a model of 
open (non-Hamiltonian) system with the special correlation. 
Note that the nonlinear evolution of statistical systems
is considered in \cite{nn1,nn2,nn3,nn4,nn5,nn6,nn8}.

%%%%%%%%%%%%%%%%%%%%%%%%%%%%%%%%%%%%%%%%%%%%%%%%%%%%%%%%%%%%%%%%%%%%%%%%%
\section{Thermodynamics Laws for non-Hamiltonian Systems}

Let us define the mean value $f(a)$ of the function
$f({\bf r},{\bf p},a)$ by the relation
\be
f(a)=\int f ({\bf r},{\bf p},a) \rho_N({\bf r}, {\bf p}, a) 
d^N{\bf r} d^N{\bf p} 
\ee
and the variation for this function by
\be
\delta_a f ({\bf r},{\bf p},a)=\sum^{n}_{k=1} 
\frac{\partial f ({\bf r},{\bf p},a)}{\partial a_k} da_k .
\ee

The first law of thermodynamic states that the internal energy $U(a)$
may change because
of heat transfer $\delta Q$, and work of thermodynamics forces
\be \label{dA0} \delta A=\sum^n_{k=1} F_k(a) d a_k . \ee
The external parameters $a=\{a_1,a_2,...a_n\}$ 
here act as generalized coordinates.
In the usual equilibrium thermodynamics the work done does 
not entirely account for the change in the internal energy. 
The internal energy also changes because of the transfer of heat, 
and so
\be dU=\delta Q- \delta A . \ee
Since thermodynamics forces $F_k(a)$ are non-potential forces
\be \label{FaFa}
\frac{\partial F_k (a)}{\partial a_l}=
\frac{\partial F_l (a)}{\partial a_k} , 
\ee
the amount of work $\delta A$ depends on the path of transition 
from one state in parameters space to another.
For this reason $\delta A$ and $\delta Q$, taken separately, 
are not total differentials.

Let us give a statistical definition of thermodynamic
forces for the non-Hamiltonian systems in the mathematical expression 
of the analog of the first thermodynamics law for the mean values. 
It would be natural to define the internal energy
as the mean value of Hamiltonian
\be 
U(a)=\int H({\bf r},{\bf p},a) \rho_N({\bf r}, {\bf p}, a) 
d^N{\bf r} d^N{\bf p} . \ee
It follows that the expression for the total differential
has the form
\[ dU(a)=\int \delta_{a} H({\bf r},{\bf p},a) 
\rho_N({\bf r}, {\bf p}, a) d^N{\bf r} d^N{\bf p}
+\int H({\bf r},{\bf p},a) 
\delta_{a}\rho_N({\bf r}, {\bf p}, a) d^N{\bf r} d^N{\bf p}. \]
Therefore
\be \label{dU} 
 dU(a)=\int \frac{\partial H({\bf r},{\bf p},a)}{\partial a_k} 
\delta a_k \rho_N({\bf r}, {\bf p}, a) d^N{\bf r} d^N{\bf p}
+\int H({\bf r},{\bf p},a) \delta_{a}\rho_N({\bf r}, {\bf p}, a) 
d^N{\bf r} d^N{\bf p} . \ee

In the first term on the right-hand side, we can use the 
definition of phase density of the thermodynamics force
\[  F^{ph}_{k}({\bf r},{\bf p},a)=-
\frac{\partial H({\bf r},{\bf p},a)}{\partial a_k} . \]
The thermodynamics force $F_k(a)$ is a mean value of 
the phase density of the thermodynamics force
\be \label{Fia}
F_{k}(a)=\int 
F^{ph}_{k}({\bf r},{\bf p},a)\rho_N({\bf r}, {\bf p}, a) 
d^N{\bf r} d^N{\bf p} .
\ee
Using this equation we can prove relation (\ref{FaFa}).

Analyzing these expressions we see that the first term on the 
right-hand side of Eq. (\ref{dU}) answers 
for the work (\ref{dA0}) of thermodynamics
forces (\ref{Fia}), whereas the amount of the heat transfer 
is given by
\be \label{dQ} \delta Q=
\int H({\bf r},{\bf p},a) \delta_{a}\rho_N({\bf r}, {\bf p}, a) 
d^N{\bf r} d^N{\bf p} . \ee
We see that the heat transfer term accounts for the change in the
internal energy due not to the work of thermodynamics forces, but
rather to change in the distribution function
cased by the external parameters $a$. 

Now let us turn our attention to the analog of the second law 
for the non-Hamiltonian systems. 

The second law of thermodynamics has the form
\be \label{SL}
\delta Q=\theta(a) dS(a) .
\ee
This implies that there exists a function of state $S(a)$
called entropy.
The function $\theta (a)$ acts as integration factor.

Let us prove that (\ref{SL}) follows from the 
statistical definition of $\delta Q$ in Eq. (\ref{dQ}).
For Eq. (\ref{dQ}), we take the distribution that is 
defined by the Hamiltonian, and show that (\ref{dQ}) can be
reduced to (\ref{SL}).

Let us assume that Eq. (\ref{rhoH}) can be solved  in the form 
\[ H=\theta (a) h(\rho_N) , \]
where $h$ depends on the distribution $\rho_N$. The function
$\theta(a)$ is a function of the parameters $a=\{a_1,a_2,...,a_n\}$. 

We rewrite (\ref{dQ}) in the equivalent form
\be \label{dQ2} \delta Q= 
\int \Bigl( \theta (a) h(\rho({\bf r},{\bf p},a ),a)
+C(a) \Bigr)
\delta_{a}\rho_N({\bf r}, {\bf p}, a) d^N{\bf r} d^N{\bf p} .
\ee
New term with $C(a)$, which is added into this equation, is equal to
zero because of the normalization condition
of the distribution function $\rho_N$:
\[ C(a) \delta_a \int \rho_N({\bf r}, {\bf p}, a) d^N{\bf r} d^N{\bf p}=
C(a) \delta_a 1=0 . \]

We can write Eq. (\ref{dQ2}) in the form
\be \label{dQ3}
\delta Q= \theta(a)\delta_{a}
\int K(\rho_N({\bf r},{\bf p},a)) d^N{\bf r} d^N{\bf p} , \ee
where the function $K=K(\rho_N)$ is defined by
\be
\frac{\partial K(\rho_N)}{\partial \rho_N}=h(\rho_N)+C(a)/\theta(a) .
\ee

We see that the expression for $\delta Q$ is integrable.
If we take $1/\theta(a)$ for the integration factor, thus 
identifying $\theta (a)$ with the analog of absolute temperature,
then, using (\ref{SL}) and (\ref{dQ3}), we can give 
the statistical  definition of entropy
\be \label{SaT} S(a)=
\int K(\rho_N({\bf r},{\bf p},a)) d^N{\bf r} d^N{\bf p} +S_0. \ee
Here, $S_0$ is the contribution to the entropy which does not depend
on the variables $a$, but may depend on the number of particles $N$
in the system.
Note that the expression for entropy is equivalent to
the mean value of phase density function
\be S^{ph}({\bf r},{\bf p},a)=K(\rho_N({\bf r},{\bf p},a))/
\rho_N({\bf r},{\bf p},a)+C(a) . \ee
$S^{ph}$ is a function of dynamic variables 
${\bf r},{\bf p}$, and the parameters $a=\{a_1,a_2,...,a_n\}$.
The number  $N$ is an arbitrary natural number since we do not
use the condition $N>>1$ or $N \rightarrow \infty$.
Note that in the usual equilibrium thermodynamics 
the function $\theta(a)$ is a mean value of kinetic energy.
In the suggested thermodynamics for the non-Hamiltonian systems 
$\theta(a)$ is the usual function of the external parameters 
$a=\{a_1,a_2,...,a_n\}$.

%%%%%%%%%%%%%%%%%%%%%%%%%%%%%%%%%%%%%%%%%%%%%%%%%%%%%%%%%%%%%%%%%%%
\section{Conclusion}

The aim of this paper is the extension of the statistical
mechanics of conservative Hamiltonian systems to 
non-Hamiltonian and dissipative systems.
In this paper, we consider a wide class of non-Hamiltonian 
statistical systems that have (canonical or 
non-canonical) distributions that are defined by Hamiltonian.
%The classical non-Hamiltonian statistical systems can have
%canonical distribution and distribution determined by Hamiltonian.
This class can be described by the non-holonomic (non-integrable)
constraint: the velocity of the elementary phase volume change 
is directly proportional to the power of non-potential forces.  
The coefficient of this proportionality is defined by Hamiltonian. 
The special constraint allows us to derive solution 
for the distribution function of the system, even in 
far-from equilibrium situation.
These distributions, which are defined by Hamiltonian, can be derived 
analytically as solutions of the Liouville equation 
for non-Hamiltonian systems. 

The suggested class of the non-Hamiltonian systems 
is characterized by the distribution functions
that are determined by the Hamiltonian. 
The constant temperature systems \cite{HLM,E,EHFML,HG,EM,Nose},
the canonical-dissipative systems \cite{SET,Eb}, 
and the Fermi-Bose classical systems \cite{Eb} 
are the special cases of suggested class of non-Hamiltonian systems.
For the non-Hamiltonian N-particle systems of this class, we can use
the analogs of the usual thermodynamics laws.  
Note that $N$ is an arbitrary natural number since we do not
use the condition $N>>1$ or $N \rightarrow \infty$.
This allows us to use the suggested class of non-Hamiltonian 
systems for the simulation schemes \cite{FS} 
for the molecular dynamics.

In the papers \cite{Tarpla1,Tarmsu,Tartf}, the quantization
of the evolution equations for non-Hamiltonian and dissipative 
systems was suggested.
Using this quantization it is easy to derive the quantum
Liouville-von Neumann equations for the N-particle statistical
operator of the non-Hamiltonian quantum system \cite{Tarkn1}.
We can derive the canonical and non-canonical 
statistical operators that are 
determined by the Hamiltonian \cite{Tarpre02,Tarpla02}.
The condition for non-Hamiltonian systems 
can be generalized by the quantization 
method suggested in \cite{Tarpla1,Tarmsu}.

%%%%%%%%%%%%%%%%%%%%%%%%%%%%%%%%%%%%%%%%%%%%%%%%%%%%%%%%%%%%%%%%%%%%%%%%%%

%%\today
%%%%%%%%%%%%%%%%%%%%%%%%%%%%%%%%%%%%%%%%%%%%%%%%%%%%%%%%%%%%%%%%

\end{document}